\documentclass[journal,twoside,web]{ieeecolor}
\usepackage{generic}
\usepackage{cite}
\usepackage{amsmath,amssymb,amsfonts}
\usepackage{graphicx}
\usepackage{textcomp}
\usepackage{epstopdf}
\usepackage{siunitx}
\usepackage{amssymb}
\usepackage{subfig}
\def\BibTeX{{\rm B\kern-.05em{\sc i\kern-.025em b}\kern-.08em
                T\kern-.1667em\lower.7ex\hbox{E}\kern-.125emX}}
\begin{document}
 
        \title{Non-Hysteretic Condition in Negative Capacitance Junctionless FETs }
 
        \author{Amin Rassekh, Farzan Jazaeri and Jean-Michel Sallese
                \thanks{Amin Rassekh, Farzan Jazaeri, and Jean-Michel Sallese are with the Electron Device Modeling and Technology Laboratory (EDLAB) of the Ecole Polytechnique F´ed´ erale de Lausanne, Switzerland (e-mail: jean-michel.sallese@epfl.ch). received XXXX XX, 2021.}}
        \maketitle
        \begin{abstract}
                This paper analyzes the design space stability of negative capacitance double gate junctionless FETs (NCDG JLFET).   Using analytical expressions derived from a charge-based model, we predict instability condition,  hysteresis voltage, and critical thickness of the ferroelectric layers giving rise to the negative capacitance behavior. The impact of the technological parameters is investigated in order to ensure hysteresis-free operation. Finally, the stability of NCDG JLFET is predicted over a wide range of temperatures from \SI{77}{K} to \SI{400}{K}. This approach has been assessed with numerical TCAD simulations.\end{abstract}
\section{Introduction}
\IEEEPARstart{A} limitation to scaling is partly due to the source/drain (S/D) junctions in conventional metal-oxide-semiconductor field-effect transistors (MOSFET). Here is where junctionless field-effect transistors (JLFET) can help to overcome this technological challenge in the nanoscale FETs \cite{lee2009junctionless}. By essence, junctionless transistors (JLTs) are free from ultra-steep junctions implantations and thermal annealing for S/D dopant activation \cite{lee2009junctionless}.
       
        Nevertheless, the scaling of MOS devices still faces the issue of power consumption and off-state current ratio \cite{takagi2007carrier}. The reduction of the power supply voltage is always the best option to mitigate power dissipation, but at the expense of I\textsubscript{on} to I\textsubscript{off} current ratio degradation since the subthreshold swing (SS) of a FET is limited to the Maxwell-Boltzmann theoretical limit, i.e.,  \SI{60}{mV/dec} at room temperature (RT).
       
        Hence, the main parameter limiting the power supply voltage scaling in FETs is subthreshold swing (SS). It was proposed that  replacing the conventional gate oxide with a stack made of an insulator and a ferroelectric material having a certain thickness could give rise to an effective negative capacitance (NC).  Negative capacitance amplifies the impact of the gate voltage seen from the channel, resulting in a rapid variation of the current.  \cite{salahuddin2008use}. This means that the SS could become lower than the Maxwell-Boltzmann theoretical limit \cite{salvatore2008demonstration,rusu2010metal,salahuddin2008use}.
       
        However, ferroelectric material can be responsible for hysteresis which is inappropriate for logic applications \cite{jo2016negative}. An NC FET should operate in a non-hysteretic regime. This happens when the total energy of series capacitors (i.e. FE + MOS) satisfies stability conditions  \cite{salahuddin2008use}.
 
The behavior of a ferroelectric material was modelled by Ginzburg and Devonshire relying on the Landau theory of phase transitions \cite{landau1937theory}.
 
       In this context and following analytical models developed in \cite{sallese2011charge, jazaeri_sallese_2018, rassekh2019modeling,rassekh2020negative}, we propose to investigate in details the negative capacitance in Double-Gate junctionless FETs (DG JLFETs). In our previous work \cite{rassekh2020negative} we studied how the ferroelectric influences the current-voltage characteristics of a DG JLFET and predicted that, although NC in the junctionless FETs does not enhance the SS in the depletion region, swing in the moderate drain currents decreases and improves the overdrive voltage.
     
  In this work, we will derive explicit relationships to predict the onset of instability, i.e. where the charge-voltage curve snaps back upon hysteresis, the critical thickness of the ferroelectric, and the hysteresis voltage.
 
\section{Merging DG JLFET with Ferroelectric}
\begin{figure}[t]
\centering
\includegraphics[width=1\columnwidth]{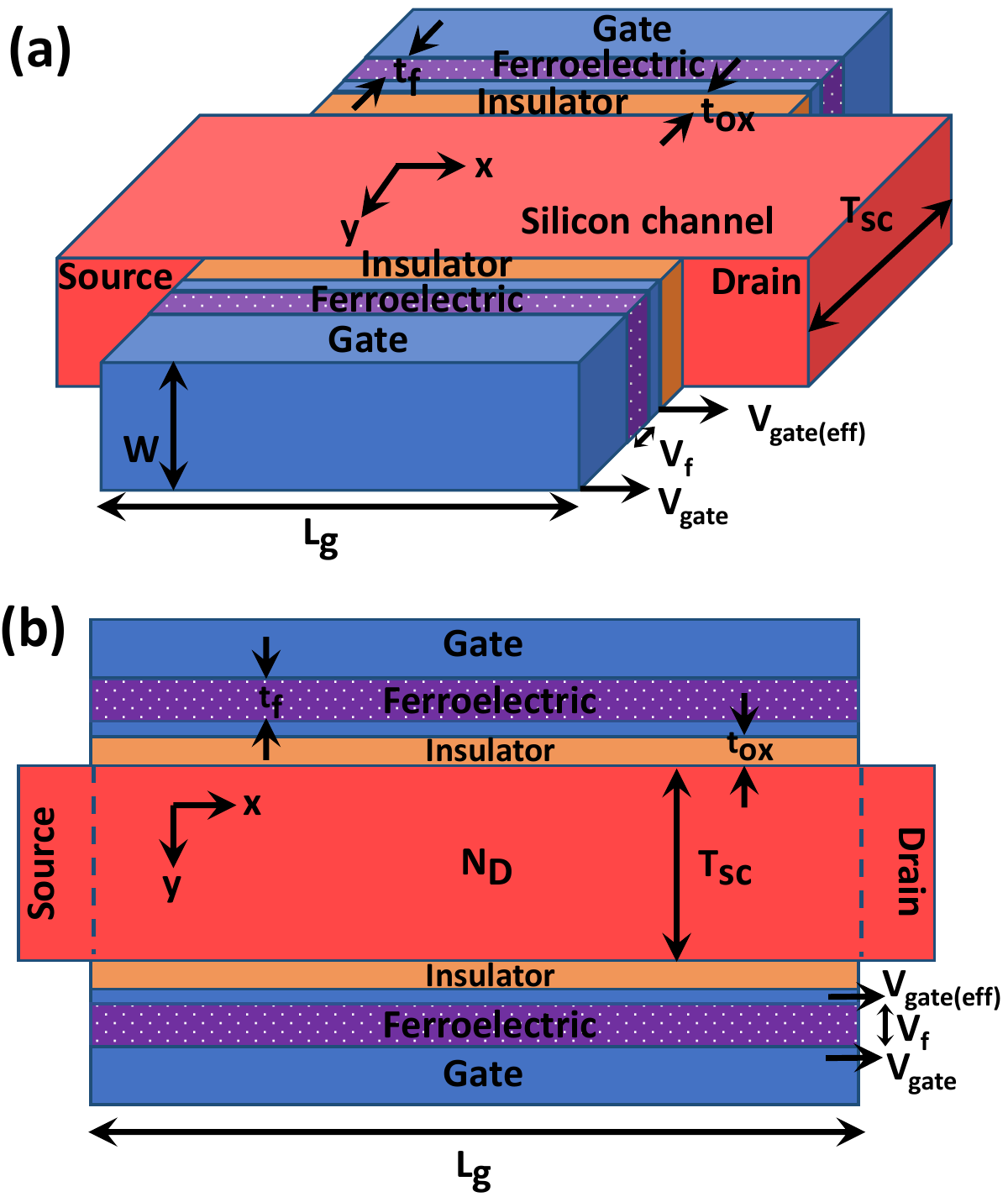}
\caption{(a) 3-D Schematic and (b) 2-D cross-section view of a symmetric  double-gate JLFET with a ferroelectric layer in the gate stack.}
\label{Fig1}
\end{figure}

        Using a ferroelectric in series with an insulating layer may, for a specific configuration, behave as a negative capacitance. If this happens, the body factor "$m$" becomes lower than unity, pushing the subthreshold swing (SS) below \SI{60}{mV/dec} \cite{salvatore2008demonstration}. To model this behavior, the common approach is to consider the Gibbs free energy $U$ of a ferroelectric with respect to the total polarization $P$  \cite{rassekh2020negative}. According to the Landau theory, the voltage drop across the ferroelectric $V_f$ is linked to the charge density of the ferroelectric $Q$ (per unit area) as follows
 
\begin{equation} \label {8}
V_f=2\alpha t_fQ+4\beta t_fQ^3+6\gamma t_fQ^5.
\end{equation}
        where $\alpha $, $\beta $ and $\gamma $ are ferroelectric material constants. The properties of a ferroelectric are also be affected by the temperature, which can be taken into account in the parameter $\alpha$, $\alpha=\alpha_0(T-T_c)$, where $T_c$ is the Curie temperature. In our paper, the Landau coefficients were extracted from the experimental data in \cite{hoffmann2019unveiling}. Three-dimensional schematics of DG JLFET structure integrated with a ferroelectric layer are drawn in Fig. \ref{Fig1}. In order to provide the uniform electric field inside the ferroelectric layer, an embedded metallic layer is introduced between the insulator and the ferroelectric \cite{rusu2010metal}. The potentials of the effective gate $V_{gate(eff)}$ and real gate $V_{gate}$ are related to the potential across the ferroelectric as follows
\begin{equation} \label {9}
V_{f}=V_{gate}-V_{gate(eff)}.
\end{equation}
\vspace{-1cm}
\subsection{DG JLFET core equations}
        An n-type long-channel symmetric double-gate JLFET with embedded ferroelectric is considered (see Fig. \ref{Fig1}). In order to focus on the negative capacitance effect of the ferroelectric in a junctionless FET, we have ignored the short channel effects intentionally. The channel length, thickness, and width are $L_g$, $T_{sc}$ and $W$ respectively. The channel doping density is $N_D$. The gate oxide and ferroelectric thicknesses are $t_{ox} $ and $t_f$ respectively. Decreasing the gate voltage in an n-type JL FET makes its channel depleted from majority carriers and puts the device in OFF-state. On the other hand, increasing the gate voltage turns on the JL FET  \cite{rassekh2020single}.
 
        According to the charge-based model for symmetric DG JLFETs proposed in \cite{sallese2011charge,jazaeri_sallese_2018}, the effective gate potential ($V_{gate(eff)}$) in depletion mode depends on the charge density in the semiconductor channel ($Q_{sc}=Q_f+Q_m$) as
\begin{equation} \label {10}
\begin{split}
V_{gate(eff)}&=\Delta\phi_{ms}+V_{ch}-\frac{Q_{sc}}{2C_{ox}}+U_{T}\ln\left(\frac{N_D}{ni}\right)\\
&+U_{T}\ln\left[1-\left(\frac{Q_{sc}}{Q_f}\right)^2\right]-\frac{Q_{sc}^2}{8C_{sc}Q_f},
\end{split}
\end{equation}
        where $n_i$ is the intrinsic carrier concentration, $C_{ox}$ and $C_{sc}=\epsilon_{si}/T_{sc}$ are the insulator and semiconductor capacitances, $U_{T}=k_BT/q$ is the thermal voltage, $V_{ch}$ is the quasi Fermi potential, $ \Delta\phi_{ms} $ is the metal - intrinsic semiconductor work function difference, $Q_m$ is the total mobile charge and $Q_f=qN_DT_{sc}$ is the fixed charge density in the channel . In accumulation mode, the effective gate voltage with respect to the charge density becomes (with $\theta=8\epsilon_{si}qN_DU_T$)
\begin{equation} \label {11}
\begin{split}
V_{gate(eff)}=\Delta\phi_{ms}+V_{ch}-&\frac{Q_{sc}}{2C_{ox}}+U_{T}\ln\left(\frac{N_D}{ni}\right)\\
&+U_{T}\ln\left(1+\frac{Q_{sc}^2}{\theta}\right).
\end{split}
\end{equation}
        Using the drift-diffusion transport model, the drain current in the channel is given by
\begin{equation} \label {111}
I_{ds}=\mu W\biggl(-Q_m\frac{d\psi_{s}}{dx}+U_T\frac{dQ_m}{dx}\biggr)=-\mu WQ_m\frac{dV_{ch}}{dx}.
\end{equation}
        where $\mu$ is the free carrier mobility that we assume constant in this work. In depletion mode, $dV_{ch}$ is obtained from (\ref{10}) whereas in accumulation relation (\ref{11}) is used instead. The drain current expressions are obtained from \cite{sallese2011charge,jazaeri_sallese_2018}.
\subsection{DG JLFET with NC}\vspace{-0.1cm}
        To calculate the potential across the ferroelectric from relation (\ref{8}), we need the total charge density in the channel. Explicit expressions of the total charge density in the ferroelectric for the JLFET operating either in depletion, accumulation, or hybrid modes obtain from ref \cite{rassekh2020negative}.
 
        The characteristics of the NCDG JLFET was analyzed and the analytical approach compared with TCAD. The 2-D electrostatics and transfer characteristics of the regular DG JLFET are simulated using TCAD SILVACO. On the other hand, the NCDG JLFET is simulated by a self-consistent solution of the charge-voltage characteristics of the ferroelectric obtained from  Landau equation and the baseline DG JLFET obtained from TCAD. \cite{rassekh2020negative}.
 
        We consider a \SI{1}{\micro m} channel length (to neglect short channel effects) and \SI{1}{nm} oxide thickness. The channel thickness is \SI{10}{nm}. The doping concentration is $N_D$=\SI{E19}{cm^{-3}}. Given that the relation between $V_{gate(eff)}$ and the ferroelectric total charge density $Q$ is known, the next step is to introduce $Q$ in (\ref{8}) and use equation (\ref{9}) to get the external gate voltage $V_{gate}$ applied to the device.
 
\begin{figure}[t]
\centering
\includegraphics[width=1\columnwidth]{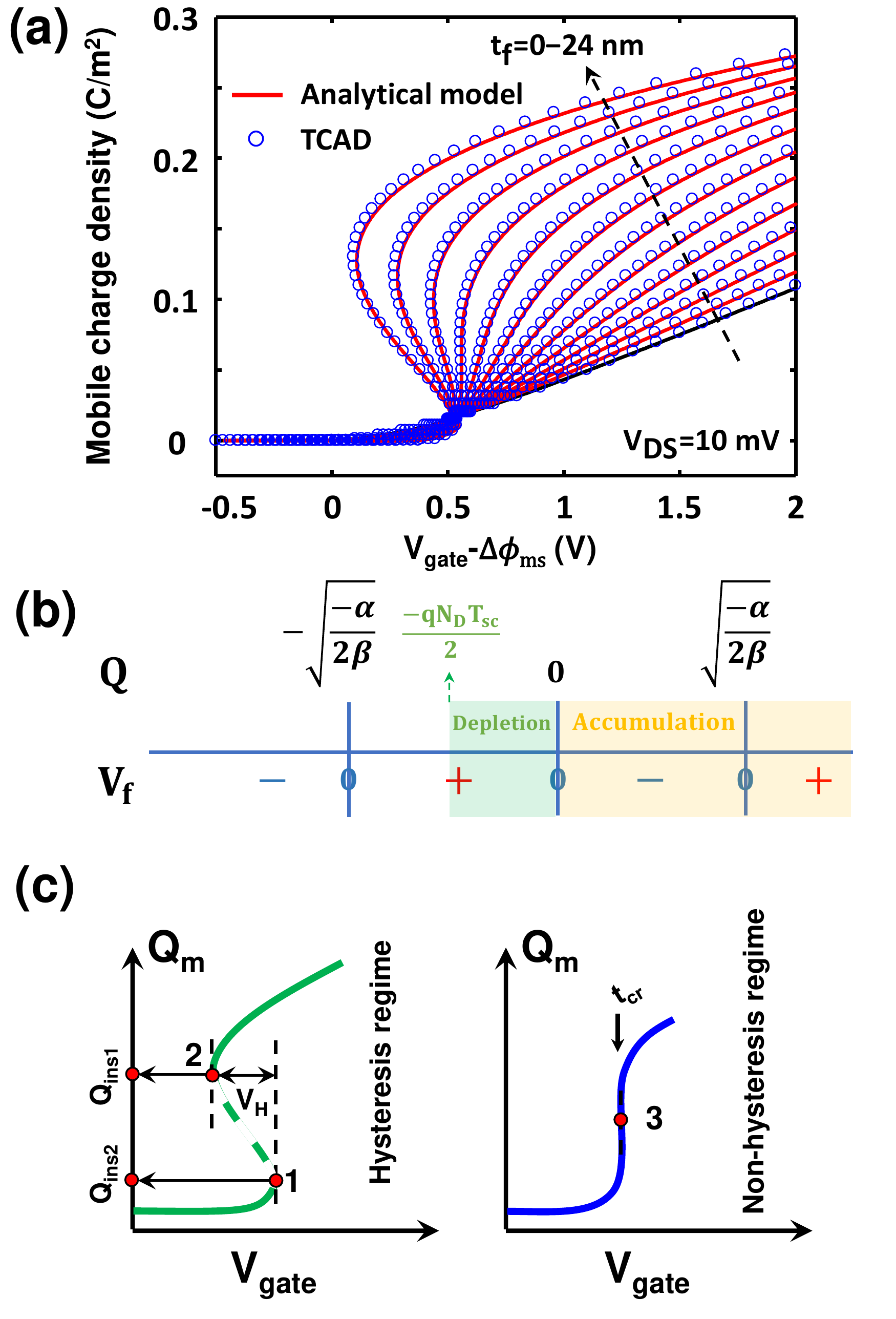}
\caption{(a)  $Q_m$ versus $V_{gate}$ from the analytical model (lines) and TCAD simulations (circles) for the various thickness of ferroelectric at  $V_{DS}$=\SI{10}{\mV}. (b) Sign table of Landau equation (we are neglecting the coefficient gamma) (c) The schematic of the curve $Q_m$-$V_{gate}$. points 1 and 2 are depicting the infinity derivative and 3 is the inflection point.}
\label{Fig2}
\end{figure}

Fig. \ref{Fig2} (a) displays the mobile charge density versus $V_{gate}$ at low $ V_{DS}$ while varying $t_f$. Fig. \ref{Fig2} (a) reveals that the slope increases when the ferroelectric thickness is increased, but above a critical value of $t_f$, that we call the critical thickness $t_{cr}$, the $Q_m-V_{gate}$ curves snap back: the ferroelectric layer starts to exhibit hysteresis. This region is where the device becomes unstable.
\section{Criteria for Instability in Linear Operation}\vspace{-0.1cm}
        As mentioned in the previous section, increasing the thickness of ferroelectric increases the slope of the mobile charge density versus gate voltage dependence, see Fig. \ref{Fig2} (a). The onset of snapback happens when the first derivative of the charge density versus the gate voltage becomes infinite. We call it the instability point. According to Fig. \ref{Fig2} (a), this condition will happen twice, see points 1 and 2 in Fig. \ref{Fig2} (c)) which is an illustration of $Q_m$-$V_{gate}$. In this case, it means that the ferroelectric will experience some hysteresis, a feature that should be avoided for a correct operation of the NC JLFET. Non-hysteresis will appear when both locus merge into a single point, i.e. point 3. This figure resumes the three important points where the curve snaps back.
 
In addition, to see voltage amplification due to the negative capacitance ,$V_f$ should be negative in relation (2). To find this condition, we first calculate the roots of relation (\ref{8}) which are given by $Q=0$, $Q=-\sqrt{\alpha/2\beta}$, and $Q=\sqrt{\alpha/2\beta}$. It is straightforward to draw the sign table of $V_f$ with respect to the roots, see Fig. \ref{Fig2} (b)).
Based on this analysis, we can already draw some important conclusion: a negative value for $V_f$ can be obtained from Eq (\ref{8}) when the total charge density in the ferroelectric material ($Q$) is either positive, or negative provided it is lower than $-\sqrt{\alpha/2\beta}$. Since this charge density is opposed to the charge in the silicon channel, $Q$=-$Q_{SC}/2$, it means that negative capacitance can always take place in accumulation mode, but not necessarily in depletion mode i.e. that NC in depletion cannot be since you need technological values ($T_{sc}$=\SI{200}{nm}, $N_{D}$=\SI{1e19}{cm^{-3}} or $T_{sc}$=\SI{10}{nm}, $N_{D}$=\SI{2e20}{cm^{-3}}), but then the JL cannot be switched off \cite{jazaeri2013modeling}. This result was not expected and is a peculiarity of junctionless FET. This means that in a junctionless FET, the NC effect in the subthreshold region, precisely before the flat band condition, can only make a shift in the $ I$-$V$ characteristics. However, in the accumulation region, or we can say above the threshold, the swing can decrease to below the \SI{60}{mV/dec}.
\subsection{Instability Points}
        Given that we have two instability points, we have the following scenarios. The first is when 'instability 1' happens in depletion and 'instability 2' occurs in accumulation. The second is when both 'instability 1 and 2' arise in accumulation. The case where both instability points happen in depletion must be rejected. Hence, we conclude that an instability point 2 cannot happen in depletion.
\begin{figure*}[t]
\centering
\includegraphics[width=2\columnwidth]{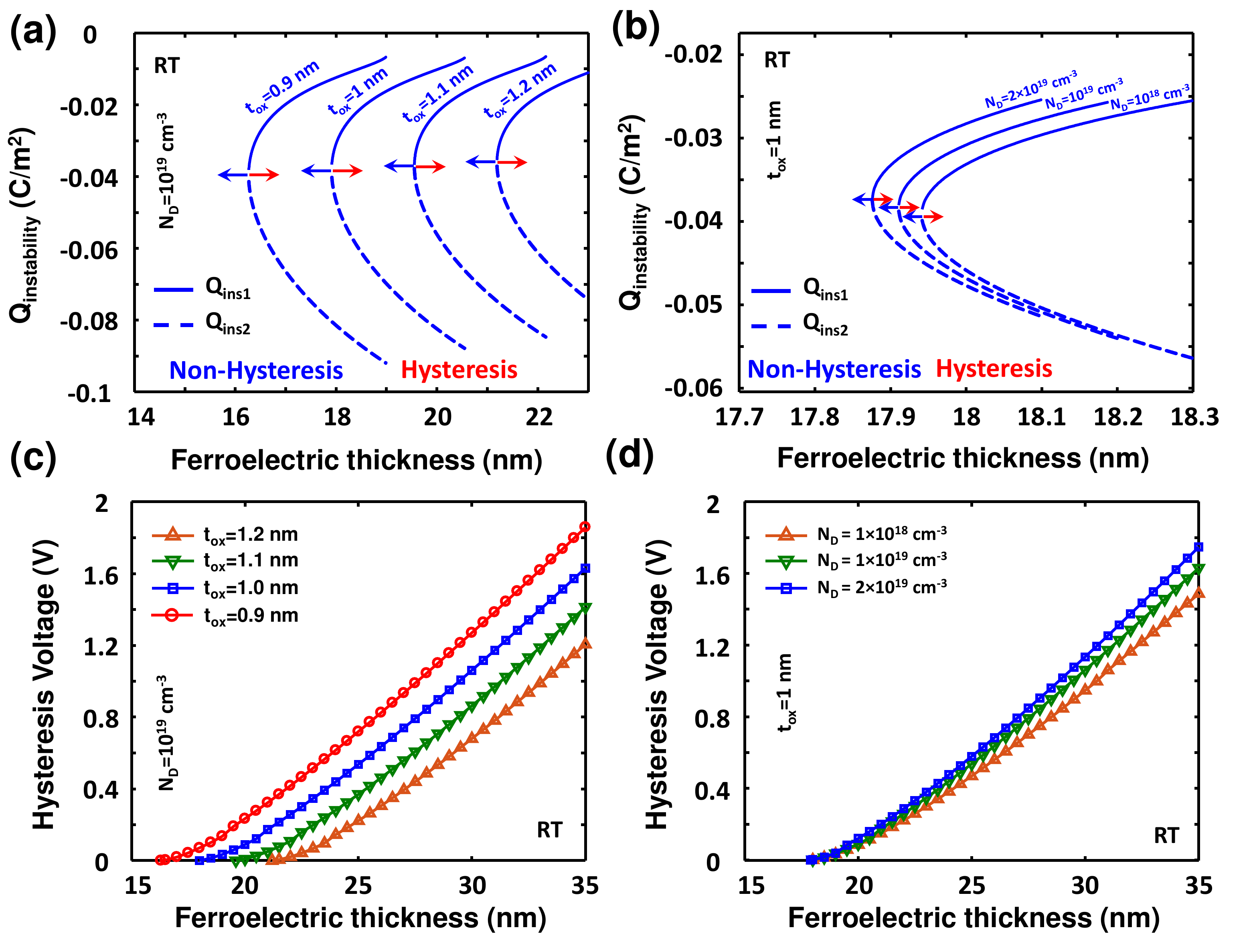}
\caption{The instability charge for (a) different oxide thickness and (b) different doping versus ferroelectric thickness. Hysteresis voltage for (c) different oxide thickness and (d) different dopingversus ferroelectric thickness.}
\label{Fig3}
\end{figure*}
\begin{figure}[t]
\centering
\includegraphics[width=1\columnwidth]{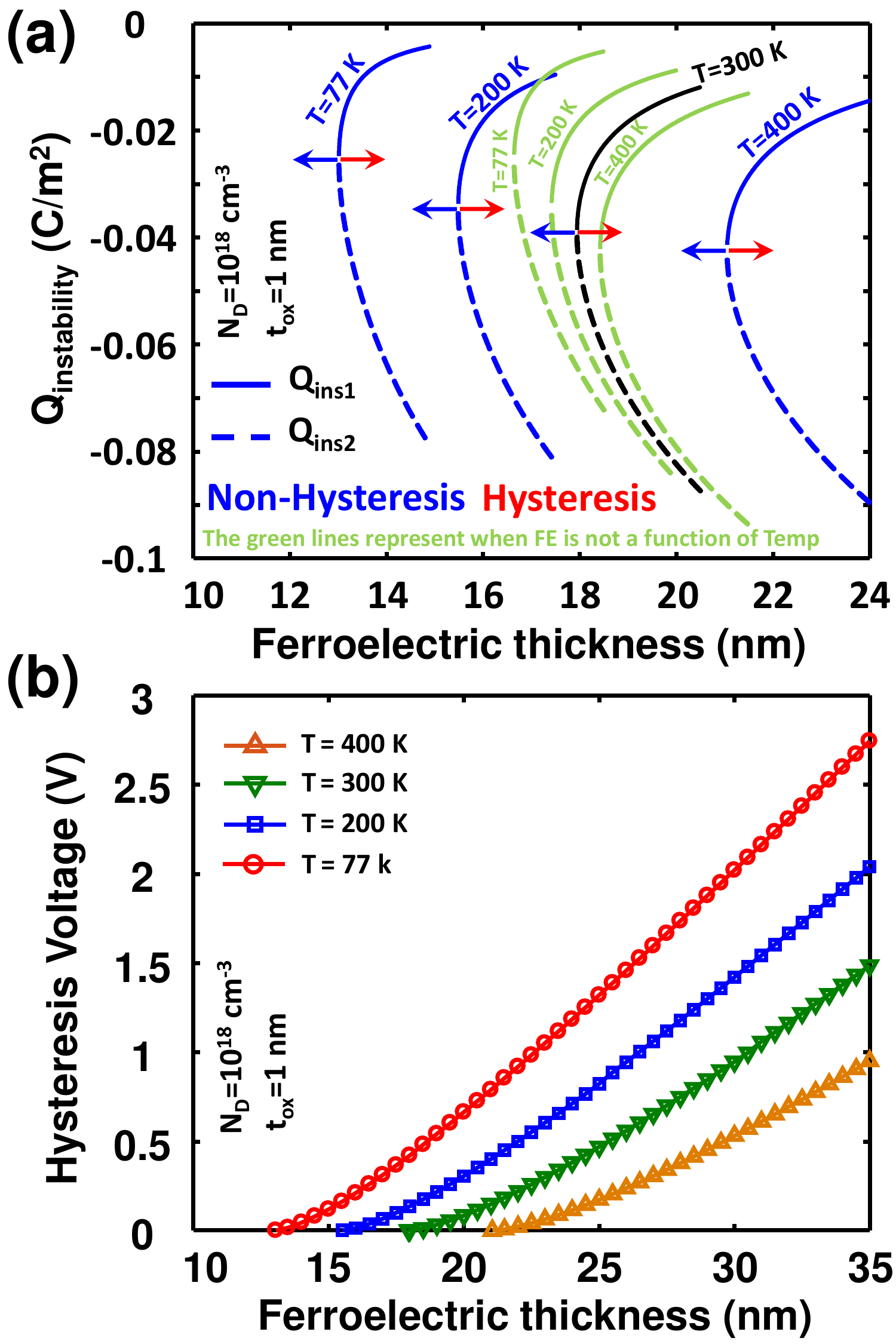}
\caption{(a) The instability charge and (b) Hysteresis voltage for the range of temperatures from \SI{77}{} to \SI{400}{\K} versus ferroelectric thickness.}
\label{Fig4}
  \end{figure}
\subsubsection{Depletion Mode}
        Calculating the instability point in depletion mode needs to get the derivative of $V_{gate}$ respect to $Q_{sc}$ from equation (\ref{10}). Introducing (\ref{8}) and (\ref{9}) in (\ref{10}) and neglecting the coefficient gamma of $V_f$ in (\ref{8}) gives
\begin{equation} \label {34}
\begin{split}
&V_{gate}-(2\alpha t_fQ+4\beta t_fQ^3)=\Delta\phi_{ms}+V_{ch}-\frac{Q_{sc}}{2C_{ox}}\\
&+U_{T}\ln\left(\frac{N_D}{ni}\right)+U_{T}\ln\left[1-\left(\frac{Q_{sc}}{Q_f}\right)^2\right]-\frac{Q_{sc}^2}{8C_{sc}Q_f},
\end{split}
\end{equation}
        where we can assume $\!Q\!\approx\!-\!Q_{sc}/2$ in linear operation. To find the instability points we must find the roots of  $dV_{gate}/dQ_{sc}$. This is given by a four degrees polynomial equation, $a_4Q_{sc}^4+a_3Q_{sc}^3+a_2Q_{sc}^2+a_1Q_{sc}+a_0=0$,
        where $a_4\!=\!-3\beta t_f/2$, $a_3\!=\!-1/(4C_{sc}Q_f)$, $a_2\!=\!3\beta t_fQ_f^2/2\!-\!\alpha t_f\!-\!1/(2C_{ox})$, $a_1\!=\!2U_T+Q_f/(4C_{sc})$ and $a_0\!=\!\alpha t_fQ_f^2+Q_f^2/(2C_{ox})$. The only acceptable roots in depletion are given by (see  \textit{Appendix I})
\begin{equation} \label {37}
Q_{scD(ins1)}=-\frac{a_3}{4a_4}+s-\frac{1}{2}\sqrt{-4s^2-2m-\frac{l}{s}},
\end{equation}
        where $m$, $l$, and $s$ are defined in \textit{Appendix I}. In addition, since the total charge density in depletion must be positive, neglecting $a_4$, $a_3$ and $a_2$ we end with a good approximation of the total charge density of the channel at the instability point.   
\subsubsection{Accumulation Mode}
        If instability happens in accumulation, we rely on the derivative of $V_{gate}$ versus $Q_{sc}$ based on (\ref{11}). Substituting (\ref{8}) and (\ref{9}) in (\ref{11}), still assuming that gamma is negligible and that in linear operation of mode $Q\!\!\approx\!\!-Q_{sc}/2$, we can write
\begin{equation} \label {39}
\begin{split}
&\!\!\!\!V_{gate}\!=\!-\alpha t_fQ_{sc}\!-\!\beta t_fQ_{sc}^3/2\!+\!\Delta\phi_{ms}\!+\!V_{ch}\!-\!\frac{Q_{sc}}{2C_{ox}}\\
&\!\!\!+U_{T}\ln\left(\frac{N_D}{ni}\right)+U_{T}\ln\left(1+\frac{Q_{sc}^2}{\theta}\right).
\end{split}
\end{equation}
        The instability point in the accumulation must satisfy $dV_{gate}/dQ_{sc}=0$, leading to
\begin{equation} \label {41}
b_4Q_{sc}^4+b_2Q_{sc}^2+b_1Q_{sc}+b_0=0,
\end{equation}
        where $b_4=-3\beta t_f/2$, $b_2=-(1/(2C_{ox})+\alpha t_f+3\beta \theta t_f/2)$, $b_1=2U_T$, and $b_0=-\theta(1/(2C_{ox})+\alpha t_f)$. Solving this equation can be done by following the same analysis as for the depletion mode (see \textit{Appendix I}). The values predicting instability in accumulation mode are
\begin{equation} \label {42}
Q_{\begin{smallmatrix}
scA(ins1)\\
scA(ins2)
\end{smallmatrix}}=-\frac{b_3}{4b_4}\mp s-\frac{1}{2}\sqrt{-4s^2-2m\pm\frac{l}{s}},
\end{equation}
        We introduce some approximation which gives the two values for the instability points in accumulation. When instability $1$ happens in accumulation, neglecting $b_4$ gives the total charge density in the channel
\begin{equation} \label {44}
Q_{scA(ins1)}\approx \frac{-b_1-\sqrt{b_1^2-4b_2b_0}}{2b_2}.
\end{equation}
        Even neglecting $b_0$, the approximation is quite accurate.
 
When instability point $1$ takes place in accumulation, it implicitly requires that the square root in (\ref{44}) must be positive, imposing a key condition on the ferroelectric thickness, $t_f<(-c_1+\sqrt{c_1^2-4c_2c_0})/2c_2$, where $c_2\!=\!3\alpha\beta\theta^2/2\!+\!\alpha^2\theta$, $c_1\!=\!3\beta\theta^2/(4C_{ox})\!+\!\alpha\theta/C_{ox}$ and $c_0\!=\!\theta/(4C_{ox}^2)\!-\!U_T^2$.
 
As mentioned, the instability point $2$ always happens in accumulation where the total charge density in the channel is negative and hence can be approximated by neglecting $b_1$ and $b_0$.
 
The instability charge of points 1 and 2 versus the thickness of the ferroelectric film for different values of oxide thickness and channel doping has been plotted in Fig. \ref{Fig3} (a), (b) respectively. As can be seen, decreasing the ferroelectric thickness causes instability points 1 and 2 to merge at point 3, and then the device operates in a non-hysteresis regime. On the other hand, increasing the oxide thickness and decreasing the channel doping lead to the non-hysteresis regime in a thicker ferroelectric film.

\subsection{Hysteresis Voltage}   
When the ferroelectric thickness is increased, the difference in terms of voltage between the instability points 1 and 2 increases. We call this difference the hysteresis voltage ($V_H$) (see Fig. \ref{Fig2} (b)). This hysteresis voltage ($V_{H}$) calculated for the NCDG JLFET is now calculated.
 
When both instability points happen in accumulation,  $V_{H}$ becomes the difference of $V_{gate}$ in two instability points $Q_{scA(ins1)}$ and $Q_{scA(ins2)}$:\\
\begin{equation} \label {A.6}
\begin{split}
&\!\!\!\!V_{H}\!=\!\big(\alpha t_f\!+\!\frac{1}{2C_{ox}}\big)(Q_{scA(ins2)}-Q_{scA(ins1)})+\\
&\!\!\!\!\frac{\beta}{2}t_f(\!Q_{scA(ins2)}^3\!-\!Q_{scA(ins1)}^3\!)\!+\!U_T\!\ln\!\bigg(\!\frac{Q_{scA(ins1)}^2\!+\!\theta}{Q_{scA(ins2)}^2\!+\!\theta}\!\bigg)\!,
\end{split}
\end{equation}
Now, when the instability 1 happens in the depletion region we have to consider  $Q_{scD(ins1)}$ and $Q_{scA(ins2)}$, which gives
\begin{equation} \label {A.7}
\begin{split}
&V_{H}\!=\!\big(\alpha t_f\!+\!\frac{1}{2C_{ox}}\big)(Q_{scA(ins2)}\!-\!Q_{scD(ins1)})\!+\!\frac{\beta}{2}t_fQ_{scA(ins2)}^3\\
&\!-\!\frac{\beta}{2}t_fQ_{scD(ins1)}^3\!+\!U_T\ln\bigg[\frac{\theta\big(Q_f^2\!-\!Q_{scD(ins1)}^2\big)}{Q_f^2\big(Q_{scA(ins2)}^2\!+\!\theta\big)}\bigg]\!-\!\frac{Q_{scD(ins1)}^2}{8C_{sc}Q_f}.
\end{split}
\end{equation}
 
Hysteresis voltage versus the thickness of ferroelectric film for different values of oxide thickness and channel doping is plotted in Fig. \ref{Fig3} (c) and (d) respectively. As can be seen, increasing $t_f$ increases the hysteresis voltage almost linearly, and for a given $t_f$ the hysteresis voltage increases when either $t_{ox}$ decreases or $N_D$ increases.
\subsection{Critical Thickness}
The critical thickness of the ferroelectric $t_{cr}$ is defined as the upper limit of the ferroelectric thickness where the device operates in the non-hysteresis regime. The $t_{cr}$ corresponds to locus 3 in Fig. \ref{Fig2} (b) where instability points 1 and 2 are equal. Since instability point 2 will happen in accumulation, $t_{cr}$ must be obtained from (\ref{42}), and \textit{should be real numbers}. Therefore, if it goes to the non-hysteresis regime, (\ref{41}) does not have real roots. According to the  \textit{Appendix I},  introducing the coefficients of (\ref{41}) in (\ref{A.2}), $t_{cr}$ is obtained from the cubic equation $d_3t_{cr}^3+d_2t_{cr}^2+d_1t_{cr}+d_0=0$
where $d_3=(\alpha +3\beta \theta/2)^3-27\alpha \beta \theta(\alpha +3\beta \theta/2)$, $d_2=3\alpha^2/(2C_{ox})-45\beta \theta(\alpha +3\beta \theta/4)/(2C_{ox})$, $d_1=3\alpha/(4C_{ox}^2)-45\beta \theta/(4C_{ox}^2)+81\beta U_T^2/2$ and $d_0=1/(2C_{ox})^3$. The real root of the cubic equation gives the critical thickness of the ferroelectric in an NCDG JLFET which is obtained from Cardano's formula    
\begin{equation} \label {A.13}
\begin{split}
t_{cr}\!\!=\!\!\sqrt[3]{\frac{-A}{2}\!+\!\sqrt{\frac{A^2}{4}\!+\!\frac{B^3}{27}}}\!+\!\sqrt[3]{\frac{-A}{2}\!-\!\sqrt{\frac{A^2}{4}\!+\!\frac{B^3}{27}}}\!\!-\!\!\frac{C}{3D},
\end{split}
\end{equation}
where $A\!=\!d_1/d_3\!-\!d_2^2/(3d_3^2)$,  $B\!=\!2d_2^3/(3d_3)^3\!-\!d_1d_2/(3d_3^2)\!+\!d_0/d_3$, $C\!=\!d_2$, $f_2$, $h_3$. In conclusion, as long as $t_f<t_{cr}$, the device works in the non-hysteresis regime.

\subsection{Temperature Effect}
The instability charge of points 1 and 2 versus the thickness of the ferroelectric film for different values of the temperature ranging from \SI{77}{} to \SI{400}{\K} (note that the semiconductor is not degenerate at \SI{77}{\K} and Boltzmann statistics are still valid has been plotted in Fig. \ref{Fig4} (a). It shows that increasing the temperature moves the non-hysteresis regime in a thicker ferroelectric film. The green lines in Fig. \ref{Fig4} (a) represent $Q_{ins1,2}$ when FE is not a function of temperature which demonstrates that the semiconductor makes little changes with temperature, and most of the temperature dependence is expected to arise from the ferroelectric layer.
 
\begin{figure}[t]
\centering
\includegraphics[width=1\columnwidth]{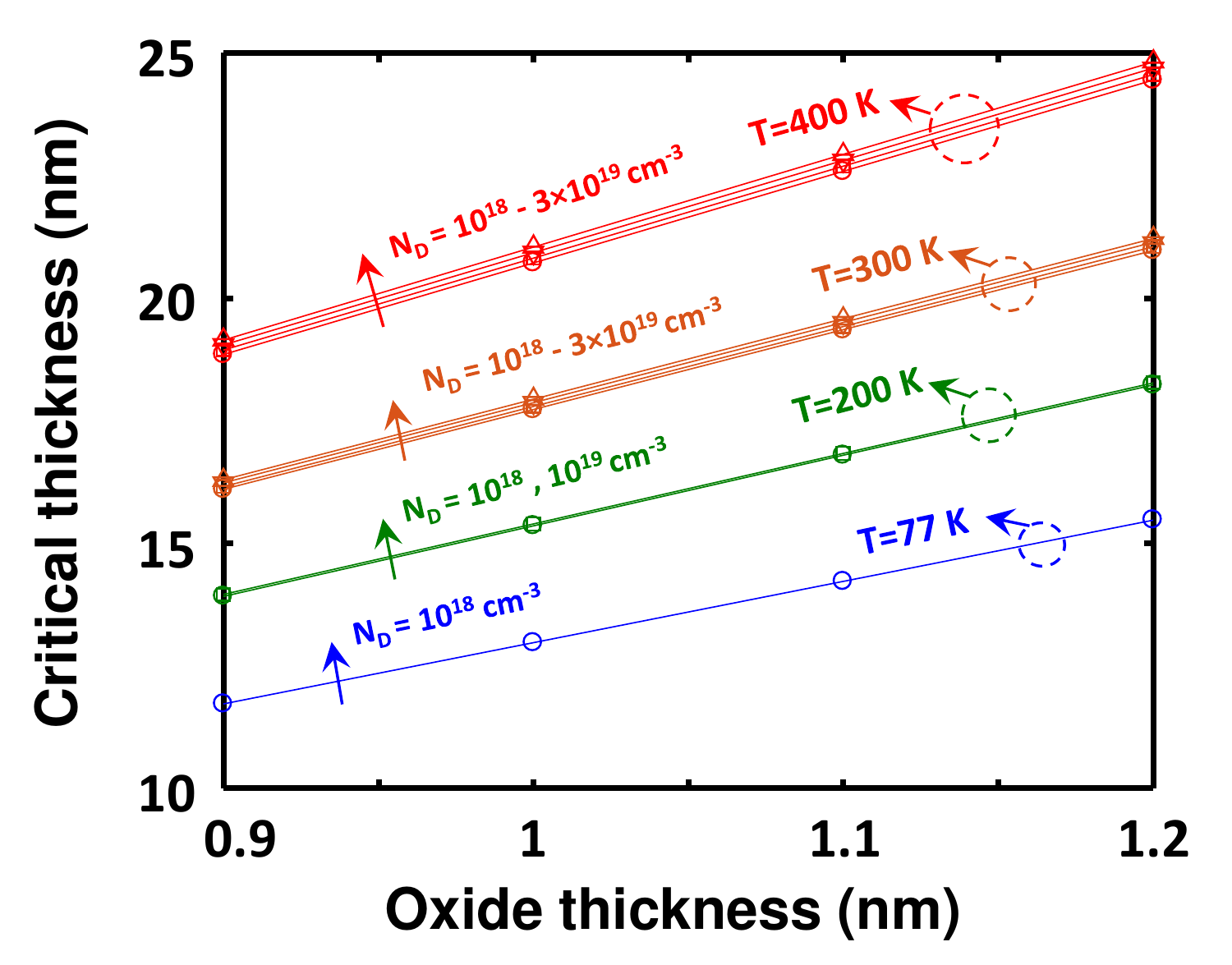}
\caption{Critical thickness versus $t_{ox}$ for different $N_D$ and temperature.}
\label{Fig5}
\end{figure}

Hysteresis voltage versus the thickness of ferroelectric film for different values of the temperature has been plotted in Fig. \ref{Fig4} (b). This figure show that the hysteresis voltage increases when temperature decreases.  
Fig. \ref{Fig5} shows $t_{cr}$ versus oxide thickness for different channel doping and different temperature. The results show that $t_{cr}$ has a linear dependency on $t_{ox}$ and is less sensitive to doping variations for lower temperatures.
\section{Conclusion}
        An analytical charge-based model for symmetric double-gate junctionless FETs with NC was developed. The model investigates the stability of an NCDG JLFET by proposing analytical expressions for the total charge density in the instability points. We proposed an explicit relationship for the critical thickness of the ferroelectric film, the thickness that determines hysteresis or non-hysteresis operation. The amount of hysteresis voltage, which is a measure of the hysteresis, has been given an analytical expression that depends explicitly on the device parameters. We also included the impact of the temperature on the JLFET with ferroelectric material from \SI{77}{\K} to \SI{400}{\K}, a significant aspect for cryogenic applications. The validity of the model has been compared to TCAD simulations with an excellent agreement in all regions of operation from linear to saturation.
        \section{APPENDIX I}
        The four roots $x_1$, $x_2$, $x_3$, and $x_4$ for a quartic equation expressed by $ax^4+bx^3+cx^2+dx+e=0$, in the case that either all roots are non-real or they are all real, are given by the following formula
        \begin{equation} \tag{A.1} \label {A.1}
        x_{1,2,3,4}=-\frac{b}{4a}-s\pm \frac{1}{2}\sqrt{-4s^2-2m\pm\frac{l}{s}},
        \end{equation}
        where $m\!=\!(8ac\!-\!3b^2)/8a^2$, $l\!=\!(b^3\!-\!4abc\!+\!8a^2d)/8a^3$ and $\!s\!=\!\!\!0.5\sqrt{-2m/3\!+\!2\Delta_0^{1/2}\cos(\phi/3)/3a}$ with $\!\!\!\phi\!\!\!=\!\arccos(\Delta_1/2\sqrt{\Delta_0^3})$, $\Delta_0\!=\!c^2\!-\!3bd\!+\!12ae$ and $\Delta_1\!=\!2c^3\!-\!9bcd\!+\!27b^2e\!+\!27ad^2\!-\!72ace$.
 
       To guarantee real roots, $s$ must be real. This happens only when the square root in $s$ becomes positive, $\Delta_0$ is positive, and $\phi$ is real. In our case, only the third one might not be met which depends on the thickness of the ferroelectric. Therefore,  following condition guarantees all real roots: $-1\leq\Delta_1/2\sqrt{\Delta_0^3}\leq 1$. The lower limit gives the critical thickness of the ferroelectric, $\Delta_1+2\sqrt{\Delta_0^3}=0$. This can be numerically solved. However, to simplify this, we assumethat $\Delta_0\approx c^2$ and $\Delta_1\approx 2c^3+27b^2e-72ace$ leading to a polynomial equation
        \begin{equation} \tag{A.2} \label {A.2}
        4c^3+27b^2e-72ace=0.
        \end{equation}

\end{document}